\documentclass{article} 
\usepackage{PRIMEarxiv}

\usepackage{longtable}
\usepackage{multirow}
\usepackage{caption} 
\usepackage{algorithm}
\usepackage{algpseudocode}
\usepackage{natbib}
\usepackage{url}
\usepackage{graphicx}
\usepackage{booktabs}
\usepackage[load-configurations=version-1]{siunitx} 

\usepackage{float}
\usepackage{titlesec}

\titlespacing*{\section}{0pt}{4pt}{3pt}
\titlespacing*{\subsection}{0pt}{4pt}{2pt}
\titlespacing*{\subsubsection}{0pt}{3pt}{1pt}

\usepackage[skip=2pt]{caption}
\title{Bridging Neural Representations and Symbolic Semantics in Audio Classification

}

\author{
Avishek Das, Pascal Hitzler \\
Kansas State University \\
Manhattan, Kansas\\
\texttt{\{avishek, hitzler\}@ksu.edu} \\
}

\begin{document}

\maketitle
\vspace{-1em}
\begin{abstract}

Deep audio classifiers can achieve strong predictive performance, but their hidden representations remain difficult to interpret. Existing audio explainability methods often focus on input-level attribution, such as waveform segments, spectrogram regions, or acoustic features, and provide limited insight into what hidden neurons encode. Interpreting hidden neurons at the concept level can provide a more meaningful view of what an audio model learns internally. This paper investigates whether hidden neurons in audio classifiers can be associated with human-understandable symbolic acoustic concepts. A model-agnostic neurosymbolic framework is presented for analyzing the final hidden layer of audio classifiers, containing 32 neurons, through symbolic concept induction. The framework is evaluated on two audio classification settings: music genre classification using GTZAN dataset and musical instrument classification using IRMAS dataset. Results show that several hidden neurons can be associated with meaningful induced concepts, and most confirmed neuron-concept pairs are statistically supported using the Mann-Whitney U test. These findings suggest that symbolic concept induction is a promising direction for explaining latent audio representations beyond input-level attribution.
\end{abstract}

\section{Introduction}
\label{sec:intro}

Deep neural networks are now being used in a wide range of audio classification systems, including speech and speaker analysis~\citep{10924161}, music information retrieval~\citep{10.1145/3714457}, bioacoustics~\citep{kumar2026applying}, environmental sound recognition~\citep{gairi2025environmental}, and clinical audio analysis~\citep{doi:10.34133/hds.0182}. However, the same depth of representation that makes these models powerful also makes them difficult to understand. Their decisions are not based on explicit, human-readable rules, but instead emerge from complex patterns of activation across time, frequency, and latent feature channels~\citep{10413597}. 
This is particularly challenging in audio, where the evidence behind a prediction can be distributed across subtle temporal and spectral patterns. In many real-world settings, such as healthcare, wildlife monitoring, accessibility, and urban safety, users need more than a predicted label; they also need to understand why the model made that prediction.

Recent reviews~\citep{akman2024audio} of explainable AI for audio suggest that the field is still less developed than comparable work in computer vision. Most existing methods continue to explain audio models through feature attribution over spectrograms~\citep{faruqui2025acoustic}, MFCCs~\citep{chakrabarty2025trust}, or waveform segments~\citep{becker2024audiomnist}, rather than through semantically meaningful descriptions of the model’s internal reasoning. This limitation is important because input-level explanations provide only a partial answer to the explainability problem. 
They can show where the model is looking, but they do not explain what a hidden unit or latent representation captures in human-understandable domain terms.
Concept-based explainability~\citep{poeta2023concept} offers a promising alternative because it shifts explanation from raw features to human-meaningful abstractions. Rather than asking which pixels, coefficients, or frames contributed most to a prediction, concept-based methods ask whether the model relied on recognizable properties such as “siren-like modulation,” “harmonic stack,” “non-rhythmic,” “bright,” or other domain-relevant semantic patterns. 
However, most of the concept-based methods are originally developed and evaluated in image-centered settings, where semantic concepts can often be grounded in visible objects, parts, textures, or scenes. Audio presents a different challenge: relevant concepts may correspond not only to identifiable sound sources, but also to temporal, spectral, rhythmic, and timbral patterns. 
For example, a neuron may respond not simply to an instrument class, but to acoustic properties such as sustained harmonic texture, percussive attack, high spectral brightness, or non-rhythmic energy distribution. These properties are meaningful to human listeners, but they are not always easy to infer from a saliency map or a highlighted spectrogram region alone.

Motivated by this challenge, this paper investigates whether hidden neurons in an audio classification model can be interpreted using human-understandable symbolic concepts. The main objectives of this paper are as follows:

\begin{itemize}
\item A neuron-level symbolic explanation pipeline is proposed for audio classifiers, combining hidden-layer activation analysis with symbolic concept induction.
\item A feature-to-tag mapping strategy is developed to convert numerical acoustic descriptors into human-understandable symbolic acoustic tags for concept induction.
\item The proposed pipeline is evaluated on both music genre and musical instrument classification tasks, demonstrating its applicability across two audio classification settings.
\item Induced neuron-level concepts are validated using independent audio data and statistical testing, providing evidence that several hidden neurons can be associated with interpretable acoustic concept expressions.
\end{itemize}


\section{Related Work}
\label{sec:related-work}

\subsection{Explainable AI for Audio}

Explainability in AI is needed for trust, transparency, debugging, and accountability, especially in high-impact domains~(\cite{gohel2021explainable}). In the audio domain,~\cite{akman2024audio} noted that many existing approaches still rely on visual or numerical explanations such as waveform relevance, spectrogram heatmaps, and feature-importance scores. Similarly, ~\citep{chakrabarty2025trust} compare LIME~\citep{ribeiro2016should}, SHAP~\citep{NIPS2017_7062}, and Grad-CAM~\citep{8237336} across speech emotion, environmental sound, and clinical audio tasks, showing that these tools can help verify whether models focus on meaningful acoustic patterns.

Several studies apply post-hoc XAI methods to specific audio tasks. \citep{becker2024audiomnist} introduce AudioMNIST and apply Layer-wise Relevance Propagation (LRP) to waveform and spectrogram models; they also propose audible heatmaps, arguing that audio explanations may be easier to interpret when presented in an auditory form. ~\citep{das2024birdsound} use LIME and SHAP for bird sound-based species recognition, while ~\citep{faruqui2025acoustic} compare LIME, SHAP, Grad-CAM, and DeepLIFT for CNN-based acoustic signal classification and show that multiple explanation methods can provide complementary evidence. In safety-critical audio settings, ~\citep{balderas2025emergency} use feature-importance explanations for emergency vehicle siren detection, and ~\citep{badawy2025sensors} use SHAP-based feature selection for vehicle fault and emergency sound classification. They both used numerical features as output for explanations. ~\citep{zinemanas2021interpretable} take a different direction by proposing an interpretable sound classification model based on learned prototypes.

These works show that most of the audio XAI methods explain predictions at the input or feature level. They can identify important spectrogram regions, waveform segments, MFCCs, or hand-crafted features, but they do not directly explain what hidden neurons or latent audio representations encode in human-understandable terms.

\subsection{Concept-Based and Neuron-Level Explainability}

Concept-based explainability shifts explanations from raw features to human-understandable concepts.~~\citep{pmlr-v80-kim18d} introduce TCAV, which tests whether user-defined concepts influence a model's prediction using concept activation vectors.~~\citep{NEURIPS2019_77d2afcb} propose ACE, which reduces manual concept specification by automatically discovering visual concepts through segmentation and clustering. Concept bottleneck models, introduced by~~\citep{pmlr-v119-koh20a}, route predictions through human-specified concept layers and allow concept-level intervention. These methods demonstrate the value of concept-level explanations, but they often require predefined concept examples, manual concept annotations, model modification, or visual segmentation procedures.

A closely related line of work focuses on interpreting hidden units directly.~~\citep{Bau_2017_CVPR} introduce Network Dissection, which measures the alignment between CNN units and semantic visual concepts.~~\citep{oikarinen2023clipdissect} propose CLIP-Dissect, which uses a vision-language model to assign open-vocabulary descriptions to neurons. These methods are powerful, but they are primarily developed for image models and depend on visual concept datasets or vision-language representations.

Neurosymbolic concept induction provides another way to generate concept-level explanations. ~\citep{dalal2024value,dalal2026toward} use positive and negative activation sets with background knowledge and ECII-based concept induction~\citep{sarker2019efficient} to assign symbolic explanations to hidden neurons in image classifiers. ~\citep{sarma2026casestudyconceptinduction} further show that this approach transfers to SUN2012, while ~\citep{das2025hidden} extend hidden-neuron activation analysis to labeled text classification using WordNet-based concepts. These studies suggest that hidden neurons can be associated with meaningful concepts, but we are not aware of any publications that explore this approach for audio.

The literature shows two main gaps. First, audio XAI is still dominated by input-level attribution methods, which explain influential input regions but not what internal units represent. Second, concept-based and neuron-level explanation methods have been developed mainly for vision and, to a lesser extent, text. This paper addresses this gap by applying neurosymbolic concept induction to hidden neurons in audio classifiers and investigating whether they can be associated with human-understandable symbolic acoustic concepts validated through activation-based statistical analysis.

\section{Methodology}
\label{method}

The methodology of this research consists of two major stages.\footnote{See \url{https://anonymous.4open.science/r/xai_audio/} for source code and raw result
files for replication.} The first stage converts raw audio samples into human-understandable symbolic tags. The second stage uses these symbolic tags, together with hidden-layer activation patterns, to induce and validate concept-level explanations for individual neurons.

\subsection{Stage 1: Audio-to-Symbolic Tag Construction}
\label{stage1}

\begin{figure}[h]
    \centering
    \includegraphics[width=0.95\textwidth]{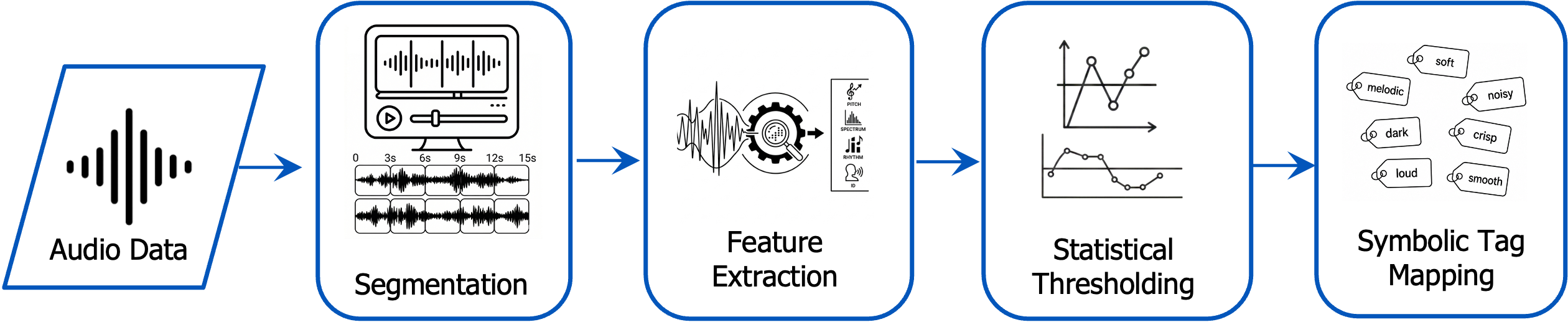}
    \caption{Audio to symbolic tag construction. Raw audio recordings are segmented, acoustic features are extracted, and statistical thresholding is used to map feature values into human-understandable symbolic tags.}
    \label{fig:symbolic-tag-generation}
\end{figure}

The goal in this stage is to bridge the gap between low-level audio signals and high-level acoustic descriptions because the concept induction module used in the later stage does not operate directly on waveforms or numerical acoustic features. Figure ~\ref{fig:symbolic-tag-generation} summarizes the tag construction steps. 

\subsubsection{Dataset}
\label{sec}

Two types of audio datasets are used in this study: 1) music genre datasets and 2) musical instrument datasets. For the song-based analysis, the GTZAN music genre dataset~\citep{tzanetakis2002musical} is used for model training and concept induction. Only five selected genre classes are considered to keep the classification setting focused and suitable for neuron-level analysis. To evaluate whether the induced concepts generalize beyond the training dataset, the FMA Large dataset~\citep{defferrard2017fma} is used for concept validation. For the instrument-based analysis, the IRMAS dataset~\citep{bosch2012comparison} is used for model training, concept induction, and concept validation. Similar to the music genre, five instrument classes are selected. This setup enables evaluation across both genre-level and instrument-level audio classification settings. Table~\ref{tab:audio_dataset_distribution} summarizes the class-wise distribution of the audio samples used in this study.  The audio files are segmented to 3-sec chunks  because they make the audio samples consistent, comparable for both model training and concept analysis. All the reported counts correspond to the 3-second audio chunks rather than full-length recordings.


\begin{table}[ht]
\small
\centering
\caption{Class-wise distribution of audio files.}
\label{tab:audio_dataset_distribution}
\begin{tabular}{lr|lr}
\hline
\multicolumn{2}{c}{\textbf{GTZAN}} & \multicolumn{2}{|c}{\textbf{IRMAS}} \\
\hline
\textbf{Class} & \textbf{Count} & \textbf{Class} & \textbf{Count} \\
\hline
Jazz      & 1000 & Electric guitar (gel)    & 760 \\
Classical & 998  & Piano (pia)              & 721 \\
Hip-hop   & 998  & Organ (org)              & 682 \\
Rock      & 998  & Acoustic guitar (gac)    & 637 \\
Country   & 997  & Cello (cel)              & 388 \\
\hline
\end{tabular}
\end{table}

\subsubsection{Feature Extraction}
\label{sec}

For each 3-second audio segment, a set of acoustic features is extracted using Librosa~\citep{mcfee2015librosa}. The extracted feature set includes chroma short-time Fourier transform statistics, root mean square energy, spectral centroid, spectral bandwidth, spectral rolloff, zero-crossing rate, harmonic and perceptual components, tempo, and Mel-frequency cepstral coefficients (MFCCs). For most descriptors, both the mean and variance are computed to capture not only the average acoustic behavior of a segment but also the variability of that behavior over time. This results in a total of 57 numerical acoustic features for each audio segment. However, numerical feature values are not directly suitable for concept induction. Therefore, these values are transformed into symbolic acoustic tags through statistical thresholding.

\subsubsection{Statistical Thresholding and Symbolic Tag Mapping}
\label{sec}

To convert numerical acoustic features into human understandable symbolic tags, a statistical thresholding is applied. Thresholds are estimated only from the training data to avoid information leakage from  test sets. For each acoustic feature, the lower and upper thresholds are defined as the 25th and 75th percentiles, respectively. 
A feature value less than or equal to the lower threshold is mapped to its corresponding low-level symbolic tag, while a value greater than or equal to the upper threshold is mapped to its corresponding high-level symbolic tag. Values falling between the two thresholds are not assigned a tag for that feature. If at least 10\% of training samples have non-zero values for a feature, thresholds are computed from the non-zero values to avoid silent chunks dominating the distribution. Otherwise, the full feature distribution is used due to insufficient non-zero samples. After the thresholds are computed, they are kept fixed and applied to the test set so that each test audio segment is mapped to its corresponding set of symbolic acoustic tags. Table~\ref{tab:symbolic_tag_mapping} summarizes the feature-to-tag mapping used for symbolic tag construction. The tag names are selected based on commonly reported relationships between acoustic descriptors and perceptual or musical properties in the audio and music information retrieval literature. The Source column lists the references used to justify each feature-tag mapping. The tags are therefore used as interpretable proxy labels for relative acoustic behavior rather than exact psychoacoustic measurements.

\begin{table*}[t]
\centering
\caption{Mapping acoustic features to symbolic tags used for concept induction. Tag names are proxy labels for relative acoustic behavior, not exact psychoacoustic measurements.}
\label{tab:symbolic_tag_mapping}
\resizebox{\textwidth}{!}{%
\begin{tabular}{p{3.5cm} | p{1.7cm} | p{1.7cm} | p{3cm} | p{5.5cm} | p{3.5cm}}
\hline
\textbf{Feature Descriptor} & \textbf{Low Tags} & \textbf{High Tags} & \textbf{Purpose} & \textbf{Tags Interpretation} & \textbf{Source} \\
\hline

Tempo (\texttt{tempo})
& Slow
& Fast
& Estimates pace in beats per minute
& Lower tempo = slower musical pacing, higher tempo = faster pacing.
& \citep{mcfee2015librosa,costa2026tempo} \\
\hline
Root Mean Square Energy (\texttt{rms\_mean})
& Soft
& Loud
& Measures average signal energy
& Lower RMS energy is treated as a proxy for softer sound, while higher RMS energy for louder sound.
& \citep{mcfee2015librosa, rawlinson2015meyda, constantinescu2023overview} \\
\hline
Spectral Centroid (\texttt{spectral\_centroid \_mean})
& Dark
& Bright
& Measures the center of mass of the spectrum
& Lower centroid refers to darker or lower frequency dominant timbre, and higher centroid to brighter timbre.
& \citep{9071125,mcfee2015librosa,rawlinson2015meyda} \\
\hline
Spectral Bandwidth (\texttt{spectral\_bandwidth \_mean})
& Pure
& Rich
& Measures the spread of spectral energy around the spectral centroid.
& Lower bandwidth = more focused spectrum, higher bandwidth = broader spectral spread and richer overtone content.
& \citep{antunes2019computer, britannica2025timbre, cusack1999effects, jannereth2021analyzing} \\
\hline
Spectral Rolloff (\texttt{rolloff\_mean})
& Mellow
& Crisp
& Measures the frequency below which most spectral energy is contained.
& Lower rolloff means most energy is concentrated in lower frequencies and interpreted as more mellow; higher rolloff indicates stronger high-frequency contribution, thus more crisp.
& \citep{10.3389/fpsyg.2022.796422, mcfee2015librosa} \\
\hline
Zero-Crossing Rate (\texttt{zero\_crossing\_rate \_mean})
& Smooth
& Noisy
& Measures how frequently the waveform changes sign over time.
& Lower zero-crossing rate is associated with smoother or more periodic signals, while higher zero-crossing rate is with noisier or more rapidly varying signals.
& \citep{gouyon2000use, essentia2026} \\
\hline
Harmonic Component Mean (\texttt{harmony\_mean})
& Non- Melodic
& Melodic
& Measures the average strength of the harmonic component 
& Lower harmonic dominance is interpreted as less melodic or less pitch-centered, while higher harmonic dominance is used as a proxy for stronger melodic or tonal content.
& \citep{6882210, 5495764} \\
\hline
Percussive Component Mean (\texttt{perceptr\_mean})
& Non- Rhythmic
& Rhythmic
& Measures the average strength of the percussive component
& Lower percussive dominance is interpreted as less rhythmically salient, while higher percussive dominance is used as a proxy for stronger rhythmic content.
& \citep{8081520, uhle2003extraction} \\

\hline
\end{tabular}%
}
\end{table*}

\subsection{Stage 2: Concept Induction Pipeline}

\begin{figure}[h]
    \centering
    \includegraphics[width=0.95\textwidth]{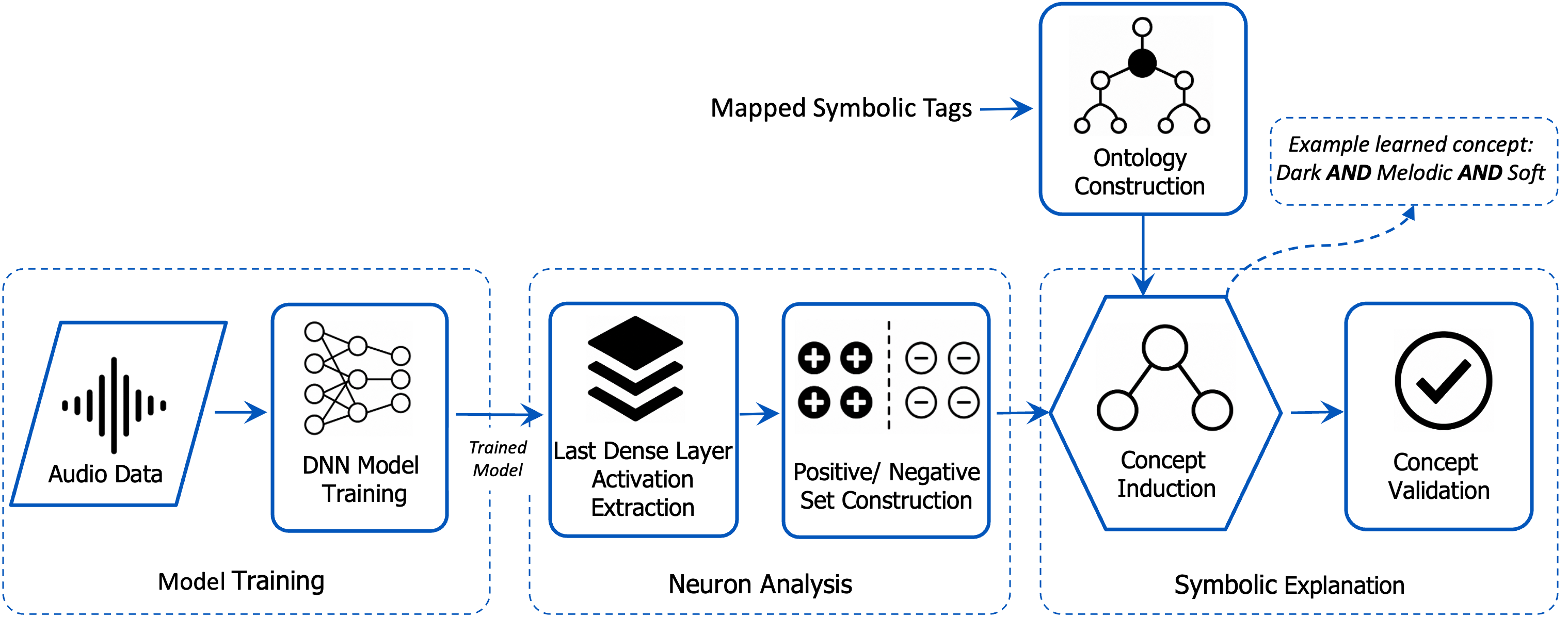}
    \caption{Overview of the proposed neuron-level symbolic explanation pipeline. A trained audio classifier is analyzed by extracting hidden-layer activations, constructing positive and negative activation sets, and applying concept induction over mapped symbolic tags to generate and validate neuron-level explanations.}
    \label{fig:neuron-level-pipeline}
\end{figure}

The second stage connects the symbolic acoustic tags generated in Stage~1 with the hidden-layer behavior of a trained audio classifier. The objective is to identify whether individual hidden neurons can be described using human-understandable acoustic concepts. The pipeline is inspired from the work by~\cite{dalal2024value} and modified accordingly for audio. Figure~\ref{fig:neuron-level-pipeline} shows the overall pipeline.

\subsubsection{Model Training}
\label{sec}
A Multi-Layer Perceptron (MLP) classifier is trained for each audio classification task using the 57 numerical acoustic features extracted from each 3-second audio segment. Both datasets are split using the 70/15/15 train/val/test ratio. The model consists of three hidden layers with 256, 128, and 32 neurons, respectively, and uses ReLU activation functions. Training is performed for 100 epochs. The final hidden layer, containing 32 neurons, is selected as the focus layer for neuron-level explanation because it provides a compact latent representation immediately before the output layer~\citep{dalal2024value, das2025hidden}.


\subsubsection{Neuron Analysis}
\label{neuron_analysis}

After classifier training, the test dataset is passed through the selected hidden layer having 32 neurons. This produces a 32-dimensional activation vector for each test audio segment, where each dimension corresponds to the response of one hidden neuron. For each neuron, positive and negative activation sets are constructed using neuron-specific thresholds. Let $a_{i,j}$ denote the activation value of neuron $j$ for audio segment i, and let $m_j = \max_i a_{i,j}$ be the maximum observed activation of neuron $j$ over the test set. The positive threshold is defined as $0.60*m_j$, and the negative threshold is defined as $0.20*m_j$. Audio segments with $a_{i,j} \geq 0.60*m_j$ are assigned to the positive set meaning strongly activating for neuron $j$, while segments with $a_{i,j} \leq 0.20*m_j$ are assigned to the negative set meaning weakly activating for that neuron. Segments with activation values between these two thresholds are treated as neutral and excluded from the concept induction problem for that neuron. The positive and negative thresholds follow the activation-discretization strategy used in prior hidden-neuron interpretation work~\citep{dalal2024value,dalal2026toward}, where examples are separated according to their activation strength relative to each neuron’s maximum response, and such cutoffs are treated as heuristic choices for obtaining strong neuron-concept associations.

\subsubsection{Concept Induction}
\label{sec}

After constructing neuron-specific positive and negative activation sets, the symbolic tags from Stage~1 (Section~\ref{stage1}) are converted into a lightweight OWL knowledge base. Each audio segment is modeled as an OWL individual, each tag as an OWL class, and tag assignments as class assertions. The resulting flat ontology contains only audio individuals and tag classes, without subclass hierarchies or additional semantic relations. For each neuron, a concept induction problem is created using its activation sets and the OWL ontology. DL-Learner~\citep{buhmann2016dl} is then applied with a closed-world reasoner and positive-negative learning formulation. The learner searches for class expressions that cover positive examples while excluding negative examples. The output is an induced concept expression for each neuron, constructed from one or more symbolic acoustic tags, which is then passed to validation.

\subsubsection{Concept validation}
\label{concept_val}
The induced concepts generated by DL-Learner are treated as neuron-level explanation hypotheses and validated using held-out audio data. For genre classification, validation samples are drawn from FMA-large; for instrument classification, a separate IRMAS subset, not used in training or concept induction is used.
For each neuron, the best induced expression divides validation samples into target and non-target groups. The target group contains samples whose symbolic acoustic tags logically satisfy the expression, while the non-target group contains samples that do not. For example, if the induced concept is ``\textit{Soft} \textbf{AND} \textit{Melodic}'', samples with both tags are assigned to the target group, and all others to the non-target group.
All validation samples are passed through the trained classifier, and activations are extracted from the same final hidden layer used in Section~\ref{neuron_analysis}. For each neuron, the positive activation threshold is computed as ($0.60 * m_j$), where $m_j$ is the maximum validation activation for neuron j. A sample is counted as a positive trigger if its activation is greater than or equal to this threshold; otherwise, it is counted as non-positive.

A neuron expression is considered confirmed if at least 50\% of the target-group samples produce positive triggers. This confirmation step checks whether samples satisfying the induced concept expression activate the corresponding neuron frequently enough to support the explanation. Confirmed neuron-concept pairs are then statistically evaluated on a separate validation split from the same source. For each confirmed neuron, activation values of target and non-target samples are compared using the non-parametric Mann-Whitney U test~\citep{mcknight2010mann}. The resulting z-score and p-value quantify whether the target group activates the neuron significantly more strongly than the non-target group.

\section{Results}
\label{result}

This section reports the results of the proposed neuron-level symbolic explanation pipeline. For the instrument classification task, the MLP achieves an overall accuracy of 0.81, with macro-average precision, recall, and F1-score all equal to 0.81. For the music genre classification task, the model achieves an overall accuracy of 0.94, with macro-average precision, recall, and F1-score all equal to 0.94. These results indicate that both classifiers learn sufficiently useful audio representations for hidden-neuron analysis. Since the objective of this study is not to optimize classifier architecture, the remaining results focus on whether neurons in the final hidden layer can be associated with human-understandable induced concepts. 
However, the next step evaluates whether the induced concepts are supported by independent validation data, as described in Section~\ref{concept_val}. Table~\ref{genre:neuron_expressions} shows the confirmed neurons for the genre classification task. The final hidden layer initially contains 32 neurons. After the label confirmation stage, 9 neurons have a positive trigger percentage greater than or equal to 50\%, and are therefore considered confirmed. The induced concept column reports the best-ranked logical expression extracted by DL-Learner for each confirmed neuron. For example, neuron\_9 is associated with audio segments described as \textit{Bright} and not \textit{Rich}, suggesting that this neuron responds to relatively bright sounds without broad spectral richness. Across the confirmed genre neurons, positive trigger percentages range from 50\% to 80\%, while negative trigger percentages remain low, with most confirmed neurons having 0\% negative triggers. 

\begin{table}[ht]
\centering
\caption{Confirmed neurons based on neuron trigger counts with best matching expressions for \textbf{Genre Classification} task. The trigger \% are based on 10 data for each neuron. Full version of the Table with all the neurons is shown in Appendix~\ref{apd:first} Table \ref{genre:neuron_expressions_full}.}
\label{genre:neuron_expressions}
\begin{tabular}{l|c|c|p{5.6cm}}
\hline
\textbf{Neuron} & \textbf{Pos. Trigger(\%)} & \textbf{Neg. Trigger(\%)} & \textbf{Induced Concepts} \\
\hline
neuron\_22 & 80 & 0 & Noisy and Soft \\
neuron\_10 & 70 & 0 & Pure and Soft \\
neuron\_12 & 70 & 10 & Crisp and Smooth \\
neuron\_31 & 60 & 0 & Pure and Soft \\
neuron\_9 & 60 & 10 & Bright and (not (Rich)) \\
neuron\_21 & 60 & 0 & Noisy and Pure \\
neuron\_15 & 50 & 0 & Smooth and Soft \\
neuron\_2 & 50 & 0 & Crisp and ((Loud and Slow) or (not (Rich))) \\
neuron\_24 & 50 & 0 & Crisp and Soft \\
\hline
\end{tabular}
\vspace{-0.1cm}
\end{table}

\begin{table}[!ht]
\centering
\caption{Confirmed neurons based on neuron trigger counts with best matching expressions for \textbf{Instrument Classification} task. The trigger\% are based on 10 data for each neuron. Full version of the Table with all the neurons is shown in Appendix~\ref{apd:first} Table \ref{instru:neuron_expressions_full}.}
\label{instru:neuron_expressions}
\begin{tabular}{l|c|c|l}
\hline
\textbf{Neuron} & \textbf{Pos. Trigger(\%)} & \textbf{Neg. Trigger(\%)} & \textbf{Induced Concepts} \\
\hline
neuron\_30 & 80 & 0 & NonMelodic and Pure \\
neuron\_7  & 80 & 20 & Rich and Soft \\
neuron\_9  & 70 & 10 & Rich and Smooth \\
neuron\_22 & 60 & 0 & Loud and Noisy \\
neuron\_15 & 60 & 10 & Crisp and Smooth \\
neuron\_0  & 50 & 0 & Bright and Loud \\
neuron\_20 & 50 & 30 & Loud and Smooth \\
neuron\_17 & 50 & 0 & Dark and Loud \\
neuron\_8  & 50 & 10 & Pure and (not (Loud)) \\
neuron\_10 & 50 & 10 & Dark and Loud \\
\hline
\end{tabular}

\end{table}

Table~\ref{instru:neuron_expressions} shows the corresponding results for the instrument classification task. The final hidden layer again contains 32 neurons; however, 4 neurons are excluded because they produce zero activation. Among the remaining 28 neurons, 10 neurons are confirmed after the label confirmation stage. Similar to the genre task, positive trigger percentages range from 50\% to 80\%, while negative trigger percentages are generally lower. These results suggest that a subset of hidden neurons in both audio classifiers can be associated with stable symbolic acoustic descriptions.

The confirmed neuron-concept pairs are further evaluated using the Mann-Whitney U test. For each confirmed neuron, the null hypothesis states that there is no significant difference between the activation values of target samples, which satisfy the induced concept expression, and non-target samples, which do not satisfy the expression. Rejecting the null hypothesis indicates that target samples activate the corresponding neuron significantly more strongly than non-target samples. Tables~\ref{genre:neuron_significance} and~\ref{instru:neuron_significance} report the corresponding p-values and z-scores for the genre and instrument classification tasks, respectively.

\begin{table}[!h]
\centering

\begin{minipage}{0.48\textwidth}
\centering
\caption{Neuron-wise statistical significance results for \textbf{Genre Classification} task. \textbf{Bold} rows indicate neurons with p-value $\geq$ 0.05.}
\label{genre:neuron_significance}
\begin{tabular}{lrr}
\hline
\textbf{Neuron} & \textbf{$p$-value} & \textbf{$z$-score} \\
\hline
neuron\_10 & 0.0003 & -3.5529 \\
neuron\_31 & 0.0110 & -2.4190 \\
neuron\_24 & 0.0010 & -3.2505 \\
neuron\_9  & 0.0009 & -3.1371 \\
\textbf{neuron\_21} & \textbf{0.0636} & -1.8898 \\
\textbf{neuron\_12} & \textbf{0.5642} & -0.5292 \\
neuron\_15 & 0.0097 & -2.5702 \\
neuron\_22 & 0.0038 & -2.8725 \\
neuron\_2  & 0.0091 & -2.4190 \\
\hline
\end{tabular}
\end{minipage}
\hfill
\begin{minipage}{0.48\textwidth}
\centering
\caption{Neuron-wise statistical significance results for \textbf{Instrument Classification} task. \textbf{Bold} rows indicate neurons with p-value $\geq$ 0.05.}
\label{instru:neuron_significance}
\begin{tabular}{lrr}
\hline
\textbf{Neuron} & \textbf{$p$-value} & \textbf{$z$-score} \\
\hline
neuron\_10 & 0.0212 & -2.2678 \\
\textbf{neuron\_17} & \textbf{0.1354} & -1.3607 \\
neuron\_9  & 0.0072 & -2.4946 \\
neuron\_15 & 0.0065 & -2.6944 \\
neuron\_7  & 0.0139 & -2.4190 \\
neuron\_20 & 0.0158 & -2.4190 \\
neuron\_0  & 0.0001 & -3.7796 \\
neuron\_8  & 0.0054 & -2.7969 \\
neuron\_30 & 0.0004 & -3.4773 \\
neuron\_22 & 0.0004 & -3.4773 \\
\hline
\end{tabular}
\end{minipage}

\end{table}

For the genre classification task, the null hypothesis is rejected for 7 of the 9 confirmed neurons at ($p < 0.05$). The null hypothesis is not rejected for neuron\_21 and neuron\_12, with ($p=0.0636$) and ($p=0.5642$), respectively. For the instrument classification task, the null hypothesis is rejected for 9 of the 10 confirmed neurons at ($p < 0.05$), while it is not rejected for neuron\_17 ($p=0.1354$). The negative z-scores indicate that the non-target group generally has lower activation values than the target group, supporting the interpretation that most confirmed concepts are associated with stronger activation of the corresponding neurons. The results validate that hidden neurons in audio classifiers can be associated with human-understandable symbolic acoustic concepts.

Table~\ref{tab:effective_labeled_comparison} provides a useful point of reference by comparing these results with related hidden-neuron analysis studies that also use concept-level interpretation of neuron activations. Although audio concepts are less directly grounded than visual objects or text terms, the percentage of effectively labeled and validated neurons remains in a comparable range, particularly for the instrument classification task.

\begin{table}[!h]
\centering
\caption{Comparison of effectively labeled neurons across hidden-neuron analysis studies.}
\label{tab:effective_labeled_comparison}
\resizebox{\columnwidth}{!}{%
\begin{tabular}{lcccc}
\hline
\textbf{Study / Task} & \textbf{Domain} & \textbf{Analyzed Neurons} & \textbf{Validated} & \textbf{\%} \\
\hline
Dalal et al.~\citep{dalal2024value} & Image & 64 & 19 & 29.7 \\
Sarma et al.~\citep{sarma2026casestudyconceptinduction} & Image & 64 & 29 & 45.3 \\
Das et al.~\citep{das2025hidden} & Text & 64 & 23 & 35.9 \\
Genre classification~(this paper) & Audio & 32 & 7 & 21.9 \\
Instrument classification~(this paper) & Audio & 32 & 9 & 28.1 \\
\hline
\end{tabular}%
}
\end{table}

This study has some limitations to be noted. First, the symbolic tags are based on thresholded acoustic features, so the quality of the induced concepts depends on the quality and interpretability of the feature-to-tag mapping. These tags are useful proxy descriptors, but they are not exact psychoacoustic measurements. Second, the ontology used in this study is a lightweight flat tag-based knowledge base; richer hierarchical or relational background knowledge may support more expressive concept induction. Third, the activation thresholds used for positive and negative set construction are heuristic, and different thresholds may produce different concept expressions. Fourth, only the final hidden layer of an MLP classifier is analyzed. Future work should examine other architectures, deeper layers, validation sets to assess the generality of the proposed approach.

\section{Conclusion and Future Work}
\label{sec:conclusion}

This paper presents a neuron-level symbolic explanation pipeline for audio classifiers. The proposed approach first maps numerical acoustic features into human-understandable symbolic acoustic tags and then uses hidden-layer activation patterns with concept induction to generate neuron-level explanations. Experiments on genre and instrument classification tasks show that several hidden neurons can be associated with meaningful induced concepts. These findings suggest that symbolic concept induction can provide a useful direction for moving audio XAI beyond input-level attribution toward interpretable analysis of latent neural representations.
Future work should examine the robustness of the proposed pipeline across additional audio datasets, model architectures, and audio representations. The current flat tag-based knowledge base can also be extended with richer semantic relations among acoustic concepts. Further work should explore alternative activation thresholds, larger validation sets, and human evaluation with audio-domain experts to assess whether the induced concepts are understandable and useful in practice.


\bibliographystyle{unsrt}
\bibliography{nesy2026-sample}

\clearpage

\appendix

\section{Confirmed Neurons Based on Positive/Negative Trigger Percentage}\label{apd:first}


\begin{table}[ht]
\centering
\caption{Neuron trigger counts with best matching expressions.}
\label{genre:neuron_expressions_full}
\begin{tabular}{l|c|c|p{5cm}}
\hline
\textbf{Neuron} & \textbf{Pos. Trigger(\%)} & \textbf{Neg. Trigger(\%)} & \textbf{Induced Concepts} \\
\hline
neuron\_22 & 80 & 0 & Noisy and Soft \\
neuron\_10 & 70 & 0 & Pure and Soft \\
neuron\_12 & 70 & 10 & Crisp and Smooth \\
neuron\_31 & 60 & 0 & Pure and Soft \\
neuron\_9 & 60 & 10 & Bright and (not (Rich)) \\
neuron\_21 & 60 & 0 & Noisy and Pure \\
neuron\_15 & 50 & 0 & Smooth and Soft \\
neuron\_2 & 50 & 0 & Crisp and ((Loud and Slow) or (not (Rich))) \\
neuron\_24 & 50 & 0 & Crisp and Soft \\
neuron\_30 & 40 & 0 & Crisp and Soft \\
neuron\_28 & 40 & 40 & Dark and Loud \\
neuron\_7 & 40 & 0 & Loud and Rich \\
neuron\_17 & 40 & 10 & Loud and Mellow and Smooth \\
neuron\_27 & 30 & 10 & Dark and Loud \\
neuron\_20 & 30 & 0 & Dark and Loud \\
neuron\_16 & 30 & 0 & Mellow and Noisy \\
neuron\_14 & 30 & 20 & Pure and (not (Mellow)) \\
neuron\_3 & 30 & 10 & Pure and (not (Mellow)) \\
neuron\_13 & 20 & 0 & Crisp and Smooth \\
neuron\_11 & 20 & 0 & Dark and Loud \\
neuron\_23 & 20 & 10 & Fast and Rich and (not (Crisp)) \\
neuron\_26 & 20 & 0 & Crisp and Smooth \\
neuron\_1 & 10 & 10 & Loud and Pure \\
neuron\_18 & 10 & 20 & Noisy and (not (Bright)) \\
neuron\_8 & 10 & 0 & Bright and (not (Crisp)) \\
neuron\_6 & 10 & 10 & Noisy and Soft \\
neuron\_5 & 10 & 20 & Pure and (not (Mellow)) \\
neuron\_29 & 10 & 0 & Rich and Soft \\
neuron\_0 & 10 & 10 & Crisp and Soft \\
neuron\_19 & 0 & 10 & Dark and Rich \\
neuron\_25 & 0 & 20 & Pure and (not (Soft)) \\
neuron\_4 & 0 & 10 & Mellow and Rich \\
\hline
\end{tabular}
\end{table}

\begin{table}[ht]
\centering
\caption{Neuron trigger counts with best matching expressions.}
\label{instru:neuron_expressions_full}
\begin{tabular}{l|c|c|p{5cm}}
\hline
\textbf{Neuron} & \textbf{Pos. Trigger(\%)} & \textbf{Neg. Trigger(\%)} & \textbf{Induced Concepts} \\
\hline
neuron\_30 & 80 & 0 & NonMelodic and Pure \\
neuron\_7  & 80 & 20 & Rich and Soft \\
neuron\_9  & 70 & 10 & Rich and Smooth \\
neuron\_22 & 60 & 0 & Loud and Noisy \\
neuron\_15 & 60 & 10 & Crisp and Smooth \\
neuron\_0  & 50 & 0 & Bright and Loud \\
neuron\_20 & 50 & 30 & Loud and Smooth \\
neuron\_17 & 50 & 0 & Dark and Loud \\
neuron\_8  & 50 & 10 & Pure and (not (Loud)) \\
neuron\_10 & 50 & 10 & Dark and Loud \\
neuron\_11 & 40 & 20 & Melodic and Soft \\
neuron\_13 & 30 & 10 & Bright and Soft \\
neuron\_1  & 30 & 0 & Mellow and Noisy \\
neuron\_14 & 30 & 0 & Soft and (not (Smooth)) \\
neuron\_12 & 30 & 20 & Melodic and Soft \\
neuron\_6  & 20 & 0 & Noisy and Soft \\
neuron\_5  & 20 & 10 & Bright and Soft \\
neuron\_4  & 20 & 10 & Bright and Soft \\
neuron\_25 & 20 & 0 & Dark and (not (Pure)) \\
neuron\_26 & 20 & 30 & Slow and Soft \\
neuron\_29 & 20 & 10 & Loud and Pure \\
neuron\_31 & 20 & 40 & Noisy and Soft \\
neuron\_16 & 10 & 0 & Mellow and Noisy \\
neuron\_19 & 10 & 10 & Noisy and NonRhythmic and Rich and (Fast or Melodic) \\
neuron\_3  & 10 & 10 & Dark and Loud \\
neuron\_18 & 0 & 40 & Dark and (not (Mellow)) \\
neuron\_27 & 0 & 30 & Crisp and Smooth \\
neuron\_2  & 0 & 10 & Mellow and Noisy \\
\hline
\end{tabular}
\end{table}



\end{document}